\documentclass[]{aa}
\usepackage{graphicx}
\usepackage{txfonts}
\begin{document}
   \title{The broad-band {\it XMM--Newton} and {\it INTEGRAL} spectra of bright type~1 Seyfert galaxies}

   \subtitle{}

   \author{
   F. \,Panessa\inst{1,2}, L. \,Bassani\inst{3}, A. \,De Rosa\inst{1}, A.J. \,Bird\inst{4}, A.J. \,Dean\inst{4}, 
   M. \,Fiocchi\inst{1}, A. \,Malizia\inst{3}, M. \,Molina\inst{4}, 
   P. \,Ubertini\inst{1}, R. \,Walter\inst{5,6}
          }

   \offprints{
   F. Panessa\\ \email{francesca.panessa@iasf-roma.inaf.it}
   }

   \institute{
   IASF-Roma/INAF, Via Fosso del Cavaliere 100, I-00133 Rome, Italy
    	    \and Instituto de F{\'\i}sica de Cantabria (CSIC-UC), Avda. de los Castros, 39005 Santander, Spain
            \and IASF-Bologna/INAF, Via P. Gobetti 101, I-40129 Bologna, Italy
	     \and School of Physics and Astronomy, University of Southampton, Southampton, SO17 1BJ, UK
	     \and INTEGRAL Science Data Centre, CH-1290 Versoix, Switzerland 
	     \and Geneva Observatory, University of Geneva, Chemin des Maillettes 51, 1290 Sauverny, Switzerland
}

%   \date{Received September 15, 1996; accepted March 16, 1997}

%   \date{Received September 15, 1996; accepted March 16, 1997}

   \abstract
%context heading (optional), leave it empty if necessary
{}
{The 0.5-150 keV broad-band spectra of a sample of nine 
   bright type 1 Seyfert galaxies are analyzed here. 
   These sources have been discovered/detected by INTEGRAL and subsequently observed with
   XMM--Newton for the first time with high sensitivity below 10 keV. The sample, although small,
   is representative of the population of type 1 AGN which are now being
   observed above 20 keV.}
{The intrinsic continuum has been modeled using three different
parameterizations: a power-law model, an exponential cut-off power-law and
an exponential cut-off power-law with a Compton reflection component.
In each model the presence of intrinsic absorption, a soft component and
emission line reprocessing features has also been tested.}
{A simple power-law model is a statistically good description
of most of the spectra presented here; an FeK line, fully and/or partial covering absorption
and a soft spectral component are detected in the majority of the sample sources. The average
photon index ($< \Gamma >$ $=$ 1.7 $\pm$ 0.2) is consistent, within errors, with the canonical spectral slope 
often observed in AGN although the photon index distribution peaks in our case 
at flat $\Gamma$ ($\sim$ 1.5) values. For four sources, we find a significantly 
improved fit when the power-law is exponentially cut-off at an energy which is 
constrained to be below $\sim$ 150 keV. The Compton reflection parameter could 
be estimated in only two objects of the sample and in both cases is found to be $R$ $>$ 1.
}
{}


\titlerunning{Broad-band spectra of type~1 Seyfert galaxies}
\authorrunning{F. Panessa et al.}

   \maketitle
%
%________________________________________________________________

\section{Introduction}

It is widely accepted that the hard X-ray emission from Active Galactic Nuclei (AGN) 
is primarily produced via unsaturated inverse Compton scattering of
UV--soft X-ray photons from the accretion disk by a corona of hot, 
probably thermal, electrons (e.g., Haardt \& Maraschi 1991, Zdziarski et al. 2000, Kawaguchi et al. 2001).
The emitted X-ray photon spectrum is best described by a power law and an exponential cut-off
at high energies. A reprocessing component is typically superimposed onto the cut-off power-law
continuum, emerging at energies $>$ 10 keV and is interpreted 
as Compton reflection of the power-law photons off thick matter
that can be the accretion disk itself and/or the torus envisaged in 
unified models. This component is typically accompanied by an iron K
fluorescence line. 

In recent years, the excellent performances of the {\it XMM-Newton}
and {\it Chandra} satellites have provided very good quality X-ray
spectra of Seyfert galaxies, allowing a proper determination of the
spectral parameters of the primary continuum and the superimposed
complex reprocessing features (see, e.g., Nandra et al. 2007, McKernan
et al. 2007, Piconcelli et al. 2005, Pounds \& Reeves 2002). However,
{\it BeppoSAX} spectra of Seyfert galaxies have shown that broad-band
observations are necessary to better constrain the slope of the
primary continuum and the value of the e-folding energy which has so
far only been properly measured in a few sources (Guainazzi et
al. 1999, Perola et al. 2002).  In addition, the determination of the
Compton reflection component $R$ (the solid angle in units of 2$\pi$
subtended by the reflecting material) is inadequate when observations
are restricted to less than 10 keV.

\begin{table*}[!htb]
\small{
\caption{\bf Sample of {\it INTEGRAL} AGN.}
\label{sample}
\begin{center}
\begin{tabular}{lrrlllll}
\hline
\hline
\multicolumn{1}{c}{Name} &
\multicolumn{1}{c}{RA} &
\multicolumn{1}{c}{Dec} &
\multicolumn{1}{c}{Class} &
\multicolumn{1}{c}{z} &
\multicolumn{1}{c}{N$_{H}$} &
\multicolumn{1}{c}{F$^{\dagger}_{20-40 keV}$} &
\multicolumn{1}{c}{Ref.}\\
\multicolumn{1}{c}{(1)} &
\multicolumn{1}{c}{(2)} &
\multicolumn{1}{c}{(3)} &
\multicolumn{1}{c}{(4)} &
\multicolumn{1}{c}{(5)} &
\multicolumn{1}{c}{(6)} &
\multicolumn{1}{c}{(7)} &
\multicolumn{1}{c}{(8)}  \\
\hline
\hline
LEDA~168563         &  73.021 &  49.546 & S1     & 0.0290    & 0.54 &  3.6$\pm$0.5 & 1    \\    
IGR~J07597-3842     & 119.923 & -38.719 & S1.2	 & 0.0400    & 0.60 &  2.2$\pm$0.2 & 2    \\
ESO~209-12          & 120.507 & -49.753 & S1.5   & 0.0405    & 0.238&  0.9$\pm$0.2 & 1    \\
Fairall~1146        & 129.620 & -36.013 & S1.5   & 0.0316    & 0.40 &  1.2$\pm$0.2 & 1    \\
4U1344-60           & 206.883 & -60.610 & S1.5   & 0.0129    & 1.07 &  4.3$\pm$0.2 & 3    \\
IGR~J16482-3036     & 252.050 & -30.590 & S1     & 0.0313    & 0.17 &  1.7$\pm$0.2 & 3   \\  
IGR~J16558-5203     & 254.010 & -52.062 & S1.2	 & 0.0540    & 0.304&  1.8$\pm$0.1 & 2    \\
IGR~J17418-1212     & 265.474 & -12.215 & S1	 & 0.0372    & 0.217&  1.4$\pm$0.2 & 4    \\
IGR~J18027-1455     & 270.685 & -14.916 & S1	 & 0.0350    & 0.496&  2.5$\pm$0.1 & 5   \\
\hline											  
\end{tabular}
\end{center}
Note: (1): Galaxy name. (2)-(3): Equatorial J2000 coordinates. (4): Optical classification. (5) Redshift. 
(6): Galactic absorption in units of 10$^{22}$ cm$^{-2}$. (7): Time average flux
expressed in units of mCrab (20-40 keV: 10 mCrab= 7.57 $\times$ 10$^{-11}$ erg cm$^{-2}$ s$^{-1}$),
Bird et al. 2007. (8): References: 1. NED position and classification; 
2. Masetti et al. 2006a; 3. Masetti et al. 2006b; 4. Torres et al. 2004; 5. Masetti et al. 2004.
}
\label{sample}
\end{table*}

The advent of the {\it INTEGRAL} and {\it Swift} satellites is significantly improving 
our knowledge of AGN spectra above 20 keV. Their all-sky hard X-ray surveys are increasing the  
statistics and demography of nearby AGN 
(Markwardt et al. 2005, Krivonos et al. 2005, Bassani et al. 2006, Beckmann et al. 2006, Sazonov et al. 2007)
and are allowing a characterization of their hard X-ray properties 
(Beckmann et al. 2006, Ajello et al. 2008, Winter et al. 2008, Molina et al. 2006, 
Malizia et al. 2007, Tueller et al. 2008, De Rosa et al. 2008).

In particular, the unique capabilities of the IBIS (Ubertini et al. 2003) instrument 
on board {\it INTEGRAL} allow the detection of sources above 20 keV
at the mCrab level with an angular resolution of 12$'$ and a typical point 
source localization accuracy of 2-3$'$. During its first few years of life, 
{\it INTEGRAL} has surveyed a large portion of the sky, detecting many Galactic/extra-galactic objects and
discovering new/unidentified sources (Bird et al. 2006, 2007, Bassani et al. 2006), many of which now identified as AGN.
Here we present the {\it XMM-Newton} and {\it INTEGRAL} broad-band spectral analysis of a sample of 9 sources 
associated with active galaxies: eight of them have been
observed for the first time below 10 keV with the high sensitivity allowed by {\it XMM-Newton}, 
despite being all very bright X-ray sources (a few mCrab level or 
2-10 $\times 10^{-11}$ erg cm$^{-2}$ s$^{-1}$ in the 20-100 keV band).
All of them have been optically classified as Seyfert galaxies of type 1-1.5 (Masetti et al. 2004, 2006a, 2006b).

The wide energy range covered by jointly fitting the {\it XMM-Newton} and {\it INTEGRAL} data up to 150 keV
offers an opportunity to characterize the spectral continuum and estimate, whenever possible, 
the exponential cut-off energy and the reflection component. 
Ideally, one would prefer to have X-ray and soft gamma-ray data
taken simultaneously, but this is now possible only through pointing observations
like those obtained by the {\it Suzaku} observatory. {\it INTEGRAL}, however,
has the advantage of having detected a number of new AGN for which {\it XMM-Newton} observations
are available. Even with the limitation of non simultaneous measurements, analysis
of broad-band data of a large sample of objects provides clues on the overall high energy properties
of type 1 Seyfert  galaxies.

The paper is organized in the following manner: Section 2 describes the sample 
and the {\it INTEGRAL} data used; Section 3 describes the {\it XMM-Newton} data reduction procedure;
Section 4 illustrates the broad-band {\it XMM-Newton} and {\it INTEGRAL} spectral fitting procedures in general
and for each individual source; Section 5 is devoted to the discussion of the different spectral components
found and their possible correlations. Finally, in Section 6 we report our conclusions.
Throughout this paper we assume a flat $\Lambda$CDM cosmology with ( $\Omega_{\rm M}$,  $\Omega_{\rm\Lambda}) = (0.3$,
0.7) and a Hubble constant of 70 km s$^{-1}$ Mpc$^{-1}$ (Bennett et al. 2003). 

\section{The Seyfert sample and the {\it INTEGRAL} data}

The sample analyzed in the present work is extracted from the Bassani et al. (2006) survey, 
updated to include a number of optical classifications obtained 
afterwards (see Bassani, Malizia and Stephen 2006). The sample includes
Seyfert 1-1.5 galaxies with a 20-100 keV flux less than 5 mCrab, for which {\it XMM-Newton} data 
were available at the time of writing. From the sample we have excluded sources already studied 
over a similar broad energy band, i.e., eight type 1-1.5 Seyfert galaxies 
listed in Bassani et al. (2006) survey are below the chosen threshold flux: four (NGC~4593,
NGC~6814, MKN~6 and MCG-6-30-15) have already been studied (Perola et al. 2002, 
Molina et al. 2006, Malizia et al. 2003a, Guainazzi
et al. 1999, respectively) and the remaining four (2E~1853.7+1534, ESO~323-G077, ESO~511-G030 
and IGR~J16119-6036) had no {\it XMM-Newton} data available or public at the time
this analysis was conducted.

All nine objects in the sample have been detected in X-rays before,
but with the exception of 4U~1344-60, the available data were of poorer quality or limited
to soft energy ranges (i.e., below 3 keV) than those presented here. None of the objects in the present sample had 
spectral information available in the soft gamma-ray range prior to this study, except 4U~1344-60 (Beckmann et al. 2006).
Overall we can conclude that the sample used in this work is representative of 
the population of type 1 AGN detected by {\it INTEGRAL} above 20 keV and consists of objects poorly studied so far.

\begin{table*}
\caption{\bf {\it XMM-Newton} Data Observation Details}
\small{
\begin{center}
\begin{tabular}{lcccccc}
\hline
\hline
\multicolumn{1}{c}{Name} &
\multicolumn{1}{c}{Obs. Date} &
\multicolumn{1}{c}{Exposure} &
\multicolumn{1}{c}{Filter}  &
\multicolumn{1}{c}{Counts/bin}  \\
\multicolumn{1}{c}{} &
\multicolumn{1}{c}{} &
\multicolumn{1}{c}{pn} &
\multicolumn{1}{c}{} \\
\multicolumn{1}{c}{(1)} &
\multicolumn{1}{c}{(2)} &
\multicolumn{1}{c}{(3)} &
\multicolumn{1}{c}{(4)} &
\multicolumn{1}{c}{(5)}\\
\hline
\hline
LEDA~168563             & 2007-02-26 & 9314  & medium	& 100 \\
IGR~J07597-3842 	& 2006-04-08 & 13260 & thin  	& 100\\  
ESO 209-12   		& 2006-03-25 & 5789  & thin  	& 20\\
FRL 1146                & 2006-04-15 & 3941  & thin	& 20\\  
FRL 1146                & 2006-12-12 & 7087  & thin     & 35\\  
4U1344-60 		& 2001-08-25 & 25000 & medium	& 35\\
IGR~J16482-3036		& 2006-03-01 & 6784/6296 & medium& 20\\
IGR~J16558-5203	        & 2006-03-01 & 6227  & medium 	& 20	\\
IGR~J17418-1212         & 2006-04-04 & 12109 & thin	& 50\\
IGR~J18027-1455         & 2006-03-25 & 16645 & thin	& 50\\
\hline
\end{tabular}
\end{center}
Notes: (1): Galaxy name. (2): Observation date.
(3): pn observation exposures except for IGR~J16482-3036 for which MOS1/MOS2 observation exposure is reported.
(4): pn filters. (5): Number of counts per bin used to rebin the spectral channels.} 
\label{table=obs_info}
\end{table*}

The {\it INTEGRAL} data presented here consist of several pointings
performed by the low-energy instrument IBIS/ISGRI ({\it INTEGRAL} Soft Gamma-Ray; Lebrun et al. 2003)
between revolution 12 and 429, i.e. the
period from launch (October 2002) to the end of April 2006. The IBIS/ISGRI images
for each available pointing were generated in various energy bands
using the ISDC off-line scientific analysis software OSA version 5.1
(see e.g., Goldwurm et al. (2003) for the ISGRI data analysis within OSA). 
Count rates at the position of the source were extracted
from individual images in order to provide light curves in various energy bands;
from these light curves average fluxes were then estimated and combined
to obtain the source spectrum (see Bird et al. 2006, 2007, for details). 

In Table~\ref{sample} we report the list of objects analyzed and the relevant information for each of them such as, 
the coordinates for epoch J2000, the optical classification, the redshift, the Galactic
column density, the 20-40 keV {\it INTEGRAL} flux as reported in Bird et al. (2007) and
the optical identification references.

For each AGN, we checked in the {\it XMM-Newton} field of view (typically a radius of 15$\arcmin$) 
for the presence of sources which might contribute to the high energy flux. 
Except for IGR~J07597-3842,
no bright sources were found. At $\sim$ 4$\arcmin$ from the nuclear position of
IGR~J07597-3842,  a bright point-like source is clearly detected in the image . Its {\it XMM-Newton} spectrum
is dominated by a strong soft component, while the 2-10 keV flux is only 6 $\times$10$^{-14}$ erg cm$^{-2}$ s$^{-1}$;
therefore we do not expect this source to contribute significantly to the {\it INTEGRAL} flux.

Finally, we remind the reader that while {\it XMM-Newton} are snap-shot observations, {\it INTEGRAL} provides
an average spectrum over many ks exposures spanning a few years period.

\section{The {\it XMM-Newton} data reduction}

All objects in the sample have been observed between March and August 2006.
The data have been processed starting from the odf files
with the {\it XMM-Newton} SAS software (version 7.0.0). 
Given its  higher sensitivity,  we use
the time average EPIC/pn spectrum for the analysis of each object, 
except for IGR~J16482-3036 for which only EPIC/MOS data are available. X-ray events 
corresponding to patterns 0-12 and 0-4 were selected from
the MOS and pn, respectively. We used the most updated calibration
files available at the time of the reduction for each source data.
Source light curves and spectra were extracted from circular regions of typically
50$\arcsec$ centered on the
source, while background products were obtained from 
off-set regions close to the source. Exposures have been
filtered for periods of high background and the effective
exposures are reported in Table~\ref{table=obs_info} as well as
the observation date, the pn filter and the number of counts per bin
used to rebin the spectral channels. Spectra were binned 
according to the luminosity of each source.
The ancillary and detector response matrices were generated using
the {\it XMM-Newton} SAS {\it arfgen} and {\it rmfgen} tasks. 

\section{The {\it XMM-Newton} and {\it INTEGRAL} spectral analysis}

The {\it XMM-Newton} and {\it INTEGRAL} data were fitted together and
analyzed using XSPEC v.12.4.0. 
Since the {\it XMM-Newton} and {\it INTEGRAL} observations 
are not simultaneous, a cross-calibration constant $C$
has been introduced in our best--fit models. This has been done
to take into account possible cross-calibration mismatches between
the two instruments or variability in the sources.
The constant was left free to vary and,
for each fit, its value is reported in the relevant Table. 
Galactic absorption is implicitly included in all spectral models; 
abundances are those of Anders \& Grevesse (1989).
The errors, lower and upper limits quoted correspond to 90\% confidence range for one
interesting parameter (i.e. $\Delta\chi^2 = 2.71$; Avni 1976).

\begin{table*}
\footnotesize{
\caption{\bf Best fit spectral parameters using an absorbed power-law, a soft thermal component and a Gaussian narrow line.}
\label{table=po}
\begin{center}
\begin{tabular}{lccccccccc}
\hline
\hline
\multicolumn{1}{c}{Name} &
\multicolumn{1}{c}{$N_{\rm H, int}$} &
\multicolumn{1}{c}{$\Gamma$} &
\multicolumn{1}{c}{$E_\alpha$} &
\multicolumn{1}{c}{$EW_\alpha$} &
\multicolumn{1}{c}{C} &
\multicolumn{1}{c}{$kT$/$\Gamma_{soft}$} &
\multicolumn{1}{c}{F$_{\rm 2-10}$} &
\multicolumn{1}{c}{F$_{\rm 20-100}$} &
\multicolumn{1}{c}{$\chi^{2}$/dof} \\
\multicolumn{1}{c}{(1)} &
\multicolumn{1}{c}{(2)} &
\multicolumn{1}{c}{(3)} &
\multicolumn{1}{c}{(4)} &
\multicolumn{1}{c}{(5)} &
\multicolumn{1}{c}{(6)} &
\multicolumn{1}{c}{(7)} &
\multicolumn{1}{c}{(8)} &
\multicolumn{1}{c}{(9)} &
\multicolumn{1}{c}{(10)} \\
\hline
\hline
LEDA~168563        & -  		    &1.68$^{+0.05}_{-0.05}$ & 6.44$^{+0.09}_{-0.11}$ & $<$52		& 0.6$^{+0.2}_{-0.2}$ & 3.8$^{+0.2}_{-0.2}$    & 1.46 & 5.6  & 507/488  \\	
IGR~J07597-3842    & -  		    &1.57$^{+0.01}_{-0.02}$ & 6.44$^{+0.04}_{-0.06}$ & 48$^{+39}_{-13}$ & 0.8$^{+0.1}_{-0.1}$ & 0.09$^{+0.01}_{-0.01}$ & 1.56 & 3.6  & 352/339  \\
ESO~209-12	   & -  		    &1.73$^{+0.02}_{-0.03}$ & 6.39$^{+0.04}_{-0.05}$ &163$^{+57}_{-62}$ & 1.2$^{+0.3}_{-0.2}$ & 0.05$^{+0.01}_{-0.01}$ & 0.83 & 1.9  & 500/508  \\
FRL~1146(AO3)      & 0.36$^{+0.05}_{-0.05}$ &1.84$^{+0.05}_{-0.05}$ & 6.39$^{+0.07}_{-0.07}$ & $<$146		& 1.1$^{+0.3}_{-0.3}$ & 0.15$^{+0.02}_{-0.01}$ & 1.15 & 2.0  & 418/421  \\
FRL~1146(AO4)$^{\dag}$& 0.28$^{+0.04}_{-0.04}$ &1.69$^{+0.04}_{-0.06}$ & 6.37$^{+0.03}_{-0.03}$ &111$^{+37}_{-38}$ & 0.7$^{+0.2}_{-0.2}$ & 0.16$^{+0.01}_{-0.01}$ & 1.21 & 2.0 & 424/438  \\
4U1344-60$^{\dag}$ & 0.98$^{+0.14}_{-0.17}$ &2.04$^{+0.08}_{-0.09}$ & 6.41$^{+0.03}_{-0.04}$ & 51$^{+20}_{-19}$ & 0.9$^{+0.1}_{-0.1}$ & -			  & 3.58 & 6.6 & 706/638 \\  
IGR~J16482-3036    & 0.09$^{+0.01}_{-0.01}$ &1.59$^{+0.02}_{-0.03}$ & 6.46$^{+0.06}_{-0.06}$ & 96$^{+32}_{-72}$ & 0.5$^{+0.1}_{-0.1}$ & -			  & 1.97 & 1.9 & 607/563  \\
IGR~J16558-5203$^{\dag}$& -	            &2.25$^{+0.03}_{-0.03}$ & 6.48$^{+0.05}_{-0.06}$ & 61$^{+50}_{-34}$ & 2.7$^{+0.5}_{-0.5}$ & -			  & 1.05 & 2.8 & 540/592  \\
IGR~J17418-1212$^{\dag}$& 0.13$^{+0.01}_{-0.01}$ &1.98$^{+0.01}_{-0.02}$ & 6.28$^{+0.04}_{-0.05}$ & 51$^{+37}_{-20}$ & 1.5$^{+0.3}_{-0.3}$ & -			  & 1.29 & 2.1 & 651/606 \\
IGR~J18027-1455$^{\dag}$& 0.41$^{+0.05}_{-0.05}$ &1.61$^{+0.04}_{-0.04}$ & 6.39$^{+0.01}_{-0.03}$ &134$^{+33}_{-34}$ & 2.9$^{+0.4}_{-0.3}$ & 0.19$^{+0.08}_{-0.07}$ & 0.61 & 4.6 & 363/316 \\ 
\hline
\end{tabular}
\end{center}
Note: (1): Galaxy name. 
(2): Intrinsic fully-covering column density in units of 10$^{22}$ cm$^{-2}$. (3): Power-law photon index. 
(4): Energy of the FeK line, in keV. (5): Equivalent width of the FeK line, in eV. (6):
Value of the {\it XMM-Newton}/{\it INTEGRAL} cross-calibration constant. (7):
Temperature of the thermal component ($kT$ in keV) or photon index of the 
soft power-law component ($\Gamma_{soft}$) in the case of LEDA~168563. 
(8-9): Model fluxes in the 2-10 keV and 20-100 keV bands 
in units of 10$^{-11}$ erg cm$^{-2}$ s$^{-1}$. (10): Chi-squared and degrees of freedom.
$\dag$: Sources with extra spectral components (see Section 4).}
\end{table*}

\begin{table*}
\footnotesize{
\caption{\bf Best fit spectral parameters using an exponential cut-off power-law model.}
\label{table=cut}
\begin{center}
\begin{tabular}{lcccc}
\hline
\hline
\multicolumn{1}{c}{Name} &
\multicolumn{1}{c}{$\Gamma$} &
\multicolumn{1}{c}{$E_{cut-off}$} &
\multicolumn{1}{c}{C} &
\multicolumn{1}{c}{$\chi^{2}$/dof} \\
\multicolumn{1}{c}{(1)} &
\multicolumn{1}{c}{(2)} &
\multicolumn{1}{c}{(3)} &
\multicolumn{1}{c}{(4)} &
\multicolumn{1}{c}{(5)} \\
\hline
\hline
LEDA~168563     & 1.59$^{+0.08}_{-0.10}$ & 73$^{+152}_{-36}$ & 0.9$^{+0.2}_{-0.3}$ & 500/487  \\   
IGR~J07597-3842 & 1.51$^{+0.03}_{-0.04}$ & 64$^{+40}_{-20}$  & 1.4$^{+0.3}_{-0.3}$ & 327/338   \\
ESO~209-12      & 1.73$^{+0.02}_{-0.03}$ & $>$ 60	     & 1.5$^{+0.5}_{-0.4}$ & 497/507   \\
FRL~1146(AO3)   & 1.81$^{+0.04}_{-0.04}$ & $>$ 45            & 1.3$^{+0.7}_{-0.5}$ & 417/420   \\
FRL~1146(AO4)   & 1.61$^{+0.09}_{-0.08}$ & $>$ 29            & 1.1$^{+0.6}_{-0.4}$ & 422/437   \\
4U1344-60       & 1.75$^{+0.18}_{-0.14}$ & $>$ 78            & 0.8$^{+0.2}_{-0.1}$ & 698/636  \\  
IGR~J16482-3036 & 1.53$^{+0.04}_{-0.04}$ & 65$^{+52}_{-24}$  & 0.7$^{+0.2}_{-0.2}$ & 590/562   \\
IGR~J16558-5203 & 2.25$^{+0.01}_{-0.01}$ & $>$ 134  	     & 2.9$^{+0.7}_{-0.3}$ & 540/591   \\
IGR~J17418-1212 & 1.97$^{+0.02}_{-0.02}$ & $>$ 170 	     & 1.6$^{+0.3}_{-0.3}$ & 651/605  \\
IGR~J18027-1455 & 1.53$^{+0.05}_{-0.05}$ & 108$^{+75}_{-32}$ & 3.6$^{+0.6}_{-0.5}$ & 341/315  \\ 
\hline
\end{tabular}
\end{center}
Note: (1): Galaxy name. 
(2): Power-law photon index. (3): Energy of the exponential cut-off in keV. (4): 
Value of the {\it XMM-Newton}/{\it INTEGRAL} cross-calibration constant.
(5): Chi-squared and degrees of freedom.}
\end{table*}

\begin{table*}
\footnotesize{
\caption{\bf Best fit spectral parameters using a pexrav model.}
\label{table=pex}
\begin{center}
\begin{tabular}{lccccc}
\hline
\hline
\multicolumn{1}{c}{Name} &
\multicolumn{1}{c}{$\Gamma$} &
\multicolumn{1}{c}{$E_{cut-off}$} &
\multicolumn{1}{c}{$R$} &
\multicolumn{1}{c}{C} &
\multicolumn{1}{c}{$\chi^{2}$/dof} \\
\multicolumn{1}{c}{(1)} &
\multicolumn{1}{c}{(2)} &
\multicolumn{1}{c}{(3)} &
\multicolumn{1}{c}{(4)} &
\multicolumn{1}{c}{(5)} &
\multicolumn{1}{c}{(6)} \\
\hline
\hline
LEDA~168563     & 1.67$^{+0.14}_{-0.17}$ & $>$ 41          & $<$ 1.69 & 0.7$^{+0.9}_{-0.3}$ & 499/486 \\
IGR~J07597-3842 & 1.48$^{+0.05}_{-0.03}$ & 52$^{+32}_{-7}$ & $<$ 0.59 & 1.5$^{+0.1}_{-0.2}$ & 328/338\\
ESO~209-12      & 1.71$^{+0.02}_{-0.03}$ & $>$ 62 	   & $<$ 0.77 & 1.5$^{+0.5}_{-0.5}$ & 497/506  \\      
FRL~1146(AO3)   & 2.13$^{+0.09}_{-0.08}$ & $>$ 48          & $<$ 2.00 & 0.7$^{+0.3}_{-0.2}$ & 407/419\\
FRL~1146(AO4)   & 1.68$^{+0.06}_{-0.06}$ & $>$ 30          & $<$ 2.63 & 0.9$^{+0.8}_{-0.4}$ & 421/436\\
4U1344-60       & 1.78$^{+0.14}_{-0.15}$ & $>$ 94          & $<$ 2.83 & 0.7$^{+0.2}_{-0.3}$ & 698/635\\
IGR~J16482-3036 & 1.62$^{+0.07}_{-0.07}$ & 87$^{+97}_{-22}$& 1.6$^{+1.5}_{-1.1}$ & 0.4$^{+0.2}_{-0.1}$ & 591/561 \\
IGR~J16558-5203 & 2.30$^{+0.04}_{-0.05}$ & -	   	   & $<$ 9.63 & 1.1$^{+1.6}_{-0.6}$ & 537/590 \\
IGR~J17418-1212 & 2.07$^{+0.05}_{-0.05}$ & -               & 2.0$^{+1.1}_{-0.9}$ & 0.9$^{+0.2}_{-0.2}$ & 637/605 \\
IGR~J18027-1455 & 1.48$^{+0.01}_{-0.01}$ & $>$ 400         & $<$0.35  & 3.8$^{+0.7}_{-0.7}$ & 338/314 \\
\hline
\end{tabular}
\end{center}
Note: (1): Galaxy name. 
(2): Power-law photon index. (3): Energy of the exponential cut-off in keV. (4): Reflection parameter.
(5): Value of the {\it XMM-Newton}/{\it INTEGRAL} cross-calibration constant. 
(6): Chi-squared and degrees of freedom.}
\end{table*}

The broad-band 0.5-150 keV {\it XMM-Newton} and {\it INTEGRAL} spectrum of each source has been
initially fitted with a power--law model absorbed by intrinsic cold absorption, plus a 
soft X-ray component and a narrow Gaussian emission (Fe) line.
A simple parameterization has been employed to model the soft component found to be present in six of  our sources:  
either a black body or a MEKAL thermal plasma model with temperature $kT$ provided a good description of 
the data, except for LEDA~168563,
where a soft power-law model ($\Gamma_{soft}$) was instead preferred.
All detected FeK$\alpha$ emission lines were found to be consistent with a narrow Gaussian profile, so that
the line width was fixed to $\sigma$ $=$ 10 eV.
In a few sources, the quality of the fit improves significantly with the introduction of
additional spectral components such as a partial covering 
absorption ({\it pcfabs} model in XSPEC) in 4U~1344-60 and IGR~J16558-5203. An extra Gaussian emission line 
is instead required in IGR~J17418-1212 and IGR~J18027-1455; finally an absorption edge was required in the case
of FRL~1146 (AO4).

In Table~\ref{table=po} we show the best--fit parameters together with the model fluxes in the 2-10 keV and 20-100 keV bands.
Although the best--fit solutions are obtained by including
the additional spectral components described above, the values of their parameters are not included in Table~\ref{table=po} but
are discussed in detail, for each source, in the next subsection.

\begin{figure*}[!ht]
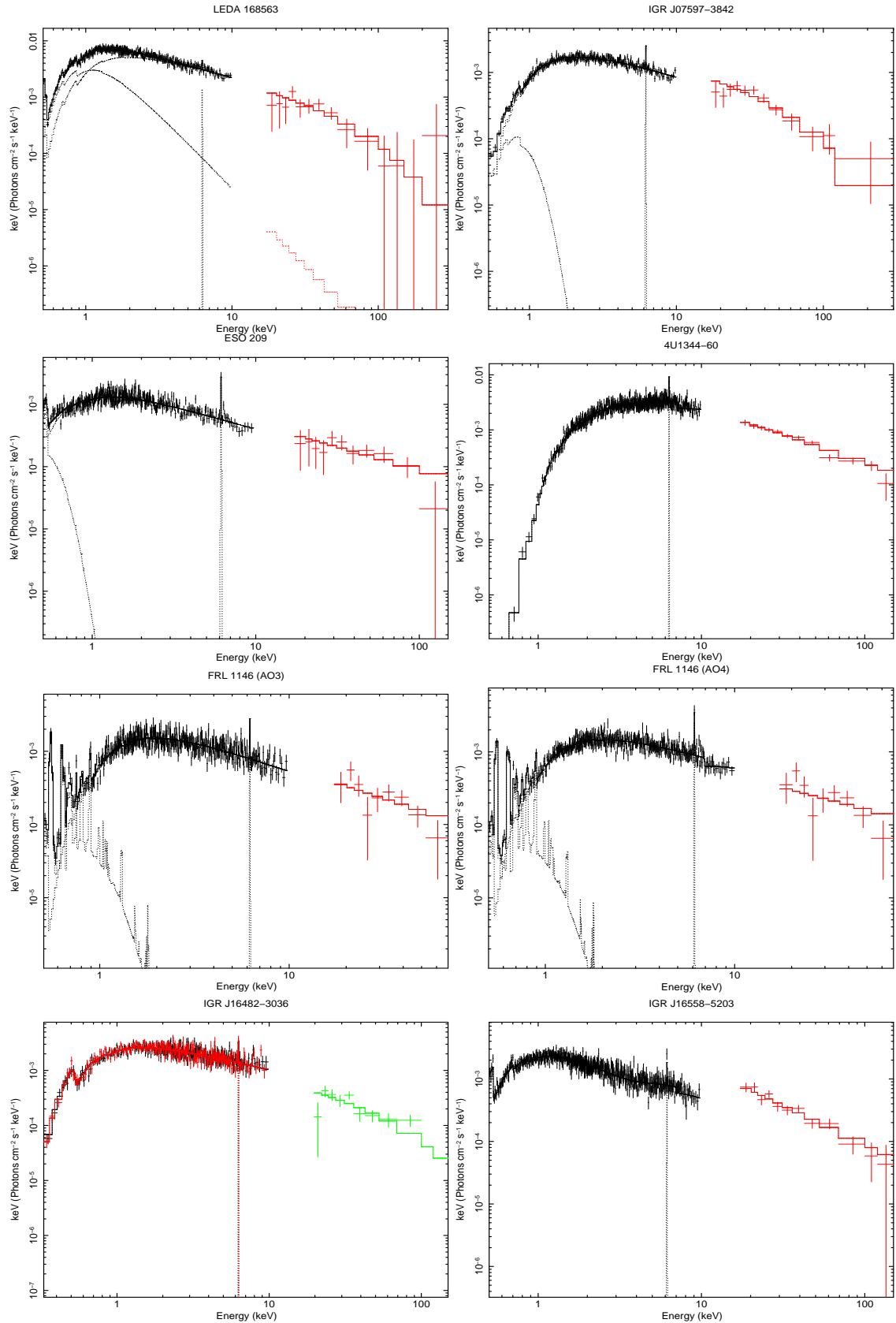

\begin{center}
\parbox{16cm}{
\includegraphics[width=0.3\textwidth,height=0.3\textheight,angle=-90]{LEDA168563_cutoff.ps}
\includegraphics[width=0.3\textwidth,height=0.3\textheight,angle=-90]{IGRJ07597_cutoff.ps}}
\parbox{16cm}{
\includegraphics[width=0.3\textwidth,height=0.3\textheight,angle=-90]{ESO209_wapo.ps}
\includegraphics[width=0.3\textwidth,height=0.3\textheight,angle=-90]{4U1344_wapo.ps}}
\parbox{16cm}{
\includegraphics[width=0.3\textwidth,height=0.3\textheight,angle=-90]{FRL1146_ao3_wapo.ps}
\includegraphics[width=0.3\textwidth,height=0.3\textheight,angle=-90]{FRL1146_ao4_wapo.ps}}
\parbox{16cm}{
\includegraphics[width=0.3\textwidth,height=0.3\textheight,angle=-90]{IGRJ16482_cutoff.ps}
\includegraphics[width=0.3\textwidth,height=0.3\textheight,angle=-90]{IGRJ16558_wapo.ps}}
\caption{The {\it XMM-Newton} and {\it INTEGRAL} best--fit spectra. Power-law model: 
ESO~209-12, FRL~1146 (AO3, AO4) and IGR~J16558-5203. Exponential cut-off power-law:
LEDA~168563, IGR~J07597-3842 and IGR~J16482-3036.}
\label{figure=bestfit1}
\end{center}
\end{figure*}

\begin{figure*}
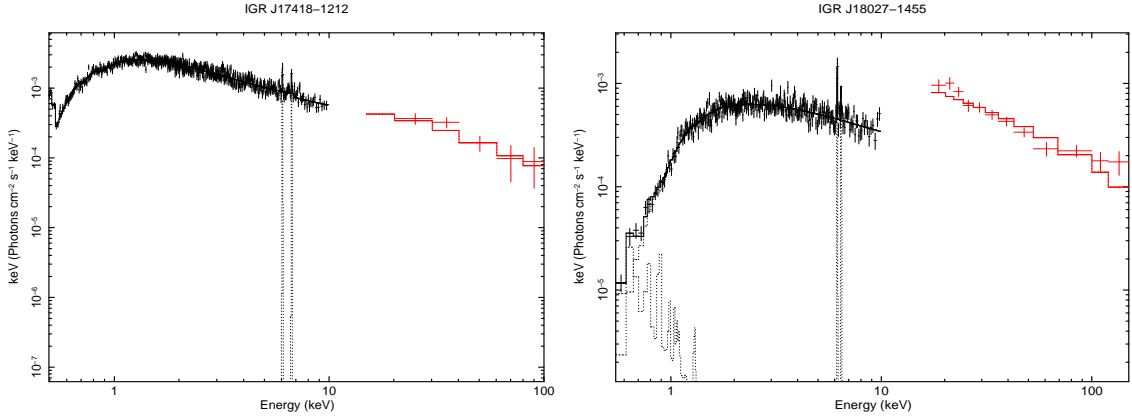

\begin{center}
\parbox{16cm}{
\includegraphics[width=0.3\textwidth,height=0.3\textheight,angle=-90]{IGRJ17418_pexrav.ps}
\includegraphics[width=0.3\textwidth,height=0.3\textheight,angle=-90]{IGRJ18027_cutoff.ps}}
\caption{Continued - The {\it XMM-Newton} and {\it INTEGRAL} best--fit spectra. Exponential cut-off power-law:
IGR~J18027-1455. Pexrav model: IGR~J17418-1212.}
\label{figure=bestfit2}
\end{center}
\end{figure*}

To check for the presence of a high energy cut-off, we replaced the simple power-law in the
best-fit model used in Table~\ref{table=po}, with
one having an exponential cut-off ({\it cutoffpl} model in XSPEC) and
report in Table~\ref{table=cut} the results provided by this change on
the main fit parameters. Finally, to also take into account the
possible presence of a reflection component, we replaced the
exponentially cut-off power-law with the {\it pexrav} model in XSPEC.
The reflection component is described by two fundamental parameters:
{\it $R$ = $\Omega$/2$\pi$}, which is the fraction of a neutral, plane
parallel slab illuminated by the power-law photons, and {\it i} or the
slab inclination angle with respect to the line of sight.  Since all
parameters turned out to be identical to those obtained 
with {\it cos i} left free to vary, we choose to freeze {\it cos i} to 0.45.  In
Table~\ref{table=pex} we report the values obtained for the {\it pexrav} main spectral parameters.

In the next subsection, we briefly describe, for each individual source, the
best--fit models and the values of the additional spectral components
which, for the sake of clarity, have not been reported in the Tables.
Note also that the values of the absorption, FeK line and soft
component parameters do not vary significantly between the power-law,
cut-off power-law and {\it pexrav} model fits, and therefore they have
not been included in Table~\ref{table=cut} and Table~\ref{table=pex}.
In Fig.~\ref{figure=bestfit1} and Fig.~\ref{figure=bestfit2}, the
unfolded spectrum and the baseline parameterization, together with the
contributions to the model of the various additive components are
shown for each object (best--fit model).

\subsection{Notes on individual sources}\label{sbys}

{\bf{LEDA~168563}.} 
The combined {\it XMM-Newton} and {\it INTEGRAL} spectrum is 
well fitted by a hard power-law model plus a soft 
power-law component with $\Gamma$ = 3.8$^{+0.2}_{-0.2}$.
A narrow Gaussian component at $\sim$ 6.4 keV is statistically significant at $\sim$ 95\% 
confidence level (equivalent width $<$ 52 eV). 
A cut-off power-law model is required by the data with a probability P $>$ 99\%, allowing 
a good constraint on the cut-off energy of $E_{cut-off}$ = 73$^{+152}_{-36}$ keV, while when using the
{\it pexrav} model, only an upper limit for the reflection parameter is obtained;
the marginal evidence for a weak iron line also suggests a low reflection component. 
Indeed, if $C$ is fixed to 1, the cut-off energy lowers slightly, while 
the upper limit on the reflection component becomes 0.6 instead of 1.7, more in line 
with the measured iron line equivalent width.
The {\it Swift}/XRT 2-10 keV flux of this source (Malizia et al. 2007)
is in good agreement with the one measured in this work and, 
together with the values of $C$, indicates that this source is not strongly variable.\\

{\bf{IGR~J07597-3842}.} 
The combined EPIC pn and {\it INTEGRAL} spectrum of this source 
is quite flat ($\Gamma$ $\sim$ 1.6) when fitted with the model
of Table~\ref{table=po}. The soft component is well represented by a black body with 
$kT$ = 0.09 $\pm$ 0.01 keV. The addition of a line
at 6.4 keV significantly improves the fit (P $>$ 99.99\%). Also a cut-off power-law model 
is highly required by the data (P $>$ 99.99\%) with a cut-off energy at 64$^{+40}_{-20}$ keV. 
Employing a {\it pexrav} model instead of the cut-off power-law does not improve the fit 
and, only yields an upper limit to the reflection parameter ($R$ $<$ 0.59).
If the cross-calibration constant is fixed to 1, the cut-off energy moves around 80 keV and the reflection becomes $\sim$ 0.4. 
The 2-10 keV flux measured by {\it Swift}/XRT (Malizia et al. 2007) is a factor of $\sim$ 1.6 higher than the
one we measure with {\it XMM-Newton}, indicating possible variability in the source. \\

{\bf{ESO~209-12}.} 
A simple power-law model for the combined {\it INTEGRAL} and
{\it XMM-Newton} spectra, plus a black body component at $kT$ = 0.05 keV
and an iron line at $\sim$ 6.4 keV provides a good fit of the spectrum.
The addition of a second Gaussian component at 6.63$^{+0.07}_{-0.08}$ keV improves the fit
with a probability P $>$ 98\%. A fit obtained using a cut-off power-law is
equally good but only gives a lower limit to the high energy cut-off ($>$ 60 keV).
Similarly, a {\it pexrav} model provides an acceptable fit to the overall source spectrum 
but no constraint on the cut-off energy, and a reflection parameter ($R$ $<$ 0.77) are obtained. 
Fixing $C$ = 1, the fit results in a flatter spectrum, a similar lower limit to the reflection, 
but a better constraint on the cut-off energy.\\

{\bf{FRL~1146}.} This source has been observed twice by {\it XMM-Newton}. The first observation
was performed on April 15th 2006 and the second on December 12th of the same year. 
No flux variability was observed between the two observations and both
spectra could be fitted with a similar spectral model, i.e., a
power-law model with intrinsic absorption (N$_{H,int}$ = 3.6$\pm$0.5
for the first observation and 2.8$\pm$0.4 $\times$ 10$^{21}$
cm$^{-2}$, for the second one), plus a MEKAL component for the soft
emission ($kT$ $\sim$ 0.15 keV in both measurements). The presence of
an iron line at $\sim$ 6.4 keV is marginally significant in the first
observation (P $\sim$ 95\%) but highly significant in the second one
(P $>$ 99.99\%). An edge at 6.93$^{+0.13}_{-0.13}$ keV with an optical
depth of 0.26$^{+0.11}_{-0.11}$ is also significantly (P $\sim$
99.95\%) detected in the second observation only. The main difference
between the two measurements is the slope of the power-law, which is
steeper in the first exposure. The cut-off energy is not properly
constrained when a cut-off power-law model is used, nor is the
reflection component in the case of the {\it pexrav} model. For both
observations $C$ is compatible with 1 within uncertainties so fixing
it to this value is acceptable, but should be regarded with caution
given the observed spectral variability: in this case the fit with
{\it pexrav} provides a cut-off energy at around 50 keV and an upper
limit for $R$ of 0.8 for the second measurement.\\

{\bf{4U~1344-60}.} The {\it XMM-Newton} data of this source have already been 
analyzed by Piconcelli et al. (2006), while the combined
{\it XMM-Newton} and {\it INTEGRAL} spectrum has been previously  
discussed by Beckmann et al. (2006) who obtained a good
representation of the broad-band data with an absorbed power-law plus a Gaussian component. 
Here, in addition to the power-law model, we used a fully-covering absorber 
(N$_{H}$=9.8 $\times$ 10$^{21}$ cm$^{-2}$) plus two layers of absorbing material partially covering
the source (N$_{H}^{pc1}$ = 4.7$^{+1.4}_{-1.2}$ $\times$ 10$^{22}$ cm$^{-2}$, CvrF$^{pc1}$ = 0.61$^{+0.06}_{-0.05}$ and
N$_{H}^{pc2}$ = 43$^{+16}_{-10}$ $\times$ 10$^{22}$ cm$^{-2}$, CvrF$^{pc2}$ = 0.50$^{+0.06}_{-0.07}$).  
The values obtained for the fully-covering absorption and for the first layer  
of partially covering material are consistent with those reported in Piconcelli
et al. (2006). However, our fit improves significantly (with a probability $>$ 99.99\%) with the addition of 
a second absorbing layer of material partially covering the AGN. A cut-off power-law model marginally 
improves the quality of the fit with a probability of $\sim$ 97\%; however, only a lower limit is found
for the high energy cut-off ($>$ 78 keV) and an upper limit to the reflection parameter ($R$ $<$ 2.8) 
if a {\it pexrav} model is used. If $C$ is fixed to 1 (which is barely allowed by the data) in the {\it pexrav} 
model, the cut-off energy is constrained to be around 140 keV and the reflection component to be $<$ 0.1 which is not
consistent with the value of the iron line equivalent width; clearly in this case a fit with a free $C$ is to be preferred.\\

{\bf{IGR~J16482-3036}.} The pn odf files of this source could not be
reprocessed forcing us to use in this case MOS1 and MOS2 data together
with the {\it INTEGRAL} data. The broad-band spectrum is well fitted by
a simple power-law model absorbed by very mild intrinsic
absorption. A narrow Gaussian component at 6.48$^{+0.08}_{-0.09}$ keV
is marginally detected (P $\sim$ 97\%); an additional narrow Gaussian
component at $\sim$ 8 keV is significant with a probability of P
$\sim$ 96\%. Possible models which could explain the presence
of an emission feature at such energies are discussed in Gallo et
al. (2005), who detected a similar feature. A cut-off
power-law model significantly (P $>$ 99.99\%) improves the quality of
the fit, allowing a constraint to be put on the high energy cut-off at
$\sim$ 65 keV.  Using the {\it pexrav} model, the fit does not
improve but provides a constrained value for the reflection component
of $\sim$ 1.6. The {\it Swift}/XRT data of this source show a spectral
slope of 1.71$^{+0.11}_{-0.12}$ plus intrinsic absorption (Malizia et
al. 2007), in perfect agreement with the {\it XMM-Newton} data when
fitted with the same model. The {\it XMM-Newton} 2-10 keV flux
is slightly higher (1.97 $\times$ 10$^{-11}$ erg cm$^{-2}$ s$^{-1}$)
than the {\it XRT} one of 1.13 $\times$ 10$^{-11}$ erg cm$^{-2}$
s$^{-1}$. In addition, the cross-calibration constant is much higher than 1,
suggesting that IGR~J16482-3036 is a variable AGN.\\

{\bf{IGR~J16558-5203}.} The spectrum of this source is well described
by a steep power-law absorbed by a partially covering medium,
significantly required with a probability $>$ 99.99\%. The column
density found for the partial absorber is 30$^{+11}_{-8}$ $\times$
10$^{22}$ cm$^{-2}$, and the covering fraction is
0.56$^{+0.06}_{-0.05}$. The presence of an iron line at 6.4 keV is
significant at only 94\% confidence level. When introduced in the fit,
the cut-off energy is constrained to be above 130 keV, but the
cross-calibration constant is quite high both in the simple and cut-off power-law
models. When the broad-band spectrum is
instead fitted with a {\it pexrav} model, $C$ is close to
1 but the reflection fraction is poorly constrained below 9.6.
If $C$ is fixed to 1 in the same model, the cut-off energy remains
unconstrained and $R$ becomes 6.6$^{+3.1}_{-2.9}$. This source
has also been observed by {\it Swift}/XRT and its power-law slope is
1.85$^{+0.06}_{-0.04}$, slightly flatter than that observed with the
pn data alone ($\Gamma$ = 2.08$\pm$ 0.01). The {\it XRT} 2-10 keV
flux is 1.77 $\times$ 10$^{-11}$ erg cm$^{-2}$ s$^{-1}$ (Malizia et
al. 2007), slightly higher than the {\it XMM-Newton} one.\\

{\bf{IGR~J17418-1212}.} The broad-band continuum of this source is
well represented by the {\it pexrav} model; fixing the cut-off
energy to a high value (which is compatible with $E_{cut-off}$ $\ge$
170 keV as indicated by the cut-off power-law fit), the reflection
parameter is constrained to be $R$ = 2.0$^{+1.1}_{-0.9}$ and the cross-calibration 
constant is found to be around 1. A first narrow Gaussian
component at 6.28$^{+0.04}_{-0.04}$ keV is significant with a 99.99\%
probability, while a second component at 6.74$^{+0.04}_{-0.06}$ keV
(equivalent widths of 63$^{+49}_{-21}$ eV) is also strongly required
by the data with a similar probability ($\sim$ 99.96\% ).\\

{\bf{IGR~J18027-1455}.} The extrapolation of the {\it XMM-Newton}
spectrum to the 20--100~keV band falls short with respect to the 
{\it INTEGRAL} detected flux; this is reflected in the relatively high
values of the cross-calibration constant ($\sim$ 3) found in our
fits. The continuum is well described by a cut-off power-law model
which allows the cut-off energy to be constrained to around 100 keV.
Only a lower limit is obtained for the reflection fraction ($R$ $<$
0.35) and a high value of the cross-calibration constant is found ($C$
$\sim$ 4). If the cross-calibration constant is fixed to 1, the
value of $R$ becomes {3.2$^{+2.2}_{-0.3}$}, suggesting a strong
reflection component, which is more compatible (but still too high)
with the equivalent width (134$^{+33}_{-34}$ eV) of the iron line at
6.4 keV. An extra narrow Gaussian component is detected
at 6.67$^{+0.06}_{-0.08}$ keV (P $\sim$ 99.80\%) with an equivalent
width of 61$^{+37}_{-28}$ eV.

\section{Discussion}

The broad-band spectra analyzed here are all well represented, to a
first approximation, by a power-law model accompanied in many cases by
extra features: a fully and/or partial covering absorption, a soft
component, one or two narrow Gaussian components and an absorption
edge. If the power-law model is replaced by a cut-off power-law
model, the fit significantly improves in the case of LEDA~168563,
IGR~J07597-3842, IGR~J16482-3036 and IGR~J18027-1455. The use of the
{\it pexrav} model allows us to constrain the reflection component
only in two objects, i.e., IGR~J16482-3036 and IGR~J17418-1212, but
only in the latter case is this model significantly required by the data.
The cross-calibration constant has been left free to vary in all our
fits and it ranges from 0.4 to 3.8, with a mean value of 1.28 in the
case of the power-law model, 1.57 for the exponential cut-off
power-law model and 1.22 when employing the {\it pexrav} model. If $C$
is fixed to 1, when this is compatible with the measured values,
better constraints on the cut-off energy and reflection are obtained
and, in most cases, these are compatible with the results reported in Tables 4
and 5. However, due to the non-simultaneity of the {\it XMM--Newton}
and {\it INTEGRAL} observations, spectral and/or flux variability
could play an important role in the determination of the spectral
parameters (as in the case of Fairall~1146 where the spectral slope
varied significantly between two different {\it XMM--Newton}
observations). In the following discussion, we choose to adopt those
fits where $C$ is left free to vary. Even in this case, some caution
is still worthwhile, but we hope that the sampling of several objects
dilutes any remaining effects of combining non-simultaneous data sets
and provides some indications on the average high energy properties of
type 1 AGN. In the following, we discuss in detail the main results
of our analysis.

\subsection{The intrinsic absorption and the soft X-ray spectrum}

In five objects of the sample, mild neutral absorption in excess of
the Galactic value is highly required to improve the quality of the fit; the
observed column densities range from 0.09 to 9.8 $\times 10^{21}$
cm$^{-2}$.  In the case of two sources, the absorption is complex
as it partially covers one source (IGR~J16558-5203), or, as in the case of
4U1344-60, is formed by two layers of material partially obscuring the central source. 
In the case of intermediate Seyfert galaxies, such as
4U1344-60, this is in agreement with the Unified Model predictions,
since at intermediate angles of the line of sight only the outer part
of the obscuring torus is intercepted by the observer (e.g., see
Maiolino 2001 and references therein). In the case of
IGR~J16558-5203, the high value of its absorption could be instead ascribed to
an intercepting cloud, a concept discussed, for instance, in Lamer et al. (2003) and
also envisaged in the clumpy torus model proposed by Elitzur \& Shlosman
(2006). In several type 1 Seyfert galaxies, complex absorption has
also been found in the form of a warm absorber (Gondoin et al. 2003,
Schurch \& Warwick 2003; Feldmeier et al. 1999), often used to
describe the soft X-ray spectrum of type 1 AGN. Since it is beyond
the scope of this work to assess the ionization state and temporal
variability of this absorber, we have chosen to describe the soft
part of the spectrum with a simple model adequate to
ensure a proper parameterization of the X-ray/gamma-ray spectrum. In
fact, a soft power-law with $\Gamma$ = 3.8 (for LEDA~168563), a black
body with temperature 0.09 and 0.05 keV (for IGR~J07597-3842 and
ESO~209-12, respectively) and a MEKAL model with $kT$ of 0.15 and 0.19
keV (for FRL~1146 and IGR~J18027-1455 respectively) provide quite good fit
to the soft energy continuum of our sources. Clearly a more detailed
analysis of the {\it XMM-Newton} high-resolution data is needed to
assess the nature of the soft component present in some of our sources
and we defer this study to a future work.

\begin{figure}[!ht]
\begin{center}
\parbox{16cm}{
\includegraphics[width=0.45\textwidth,height=0.3\textheight]{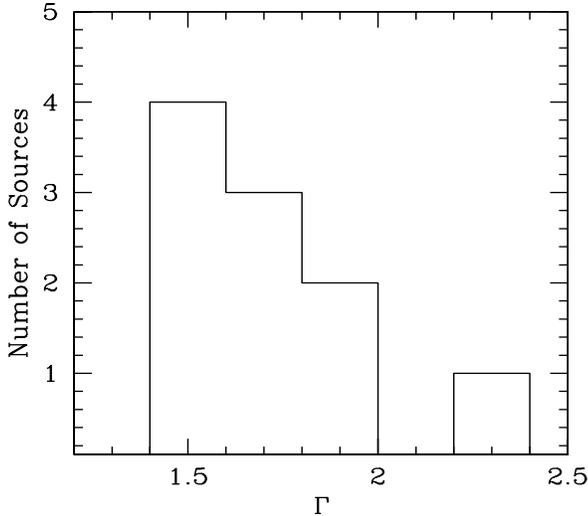}}
\caption{Photon index distribution (data from Table~\ref{table=cut}). A binning of 0.2 in photon index has been used.}
\label{figure=gamma}
\end{center}
\end{figure}

\subsection{The intrinsic primary continuum and the high energy cut-off}

In Fig.~\ref{figure=gamma} we show the photon index distribution 
obtained from data in Table~\ref{table=cut}, where a cut-off power-law model is used.
The mean $\Gamma$ is 1.73 with a standard deviation of 0.24. This value is consistent, 
within errors, with the 2-10 keV mean slope found for other sample of AGN 
(i.e., $< \Gamma >$ $\sim$ 1.8-2, Reeves \& Turner 2000, Piconcelli et al. 2005, Dadina 2008).
Four sources (LEDA~168563, IGR~J07597-3842, IGRJ16482-3036 and IGRJ18027-1455) in our sample 
have very flat spectra, i.e., $\Gamma$ $\sim$ 1.5-1.6,
providing the low $\Gamma$ peak seen in Fig.~\ref{figure=gamma}.
Our photon index distribution is similar to the one reported for
a sample of nearby Seyfert 1-1.5 galaxies by Cappi et al. (2006) who found an even flatter
weighted mean value for $\Gamma$ ($1.56 \pm 0.04$ compared to our $1.68 \pm 0.02$).
Interestingly, flat photon indices have been invoked to account for the X-ray background
spectral shape in synthesis models recently proposed (Gilli et al. 2007).
There are several examples in the literature of type 1 sources with 
flat spectra (Mkn~841, Petrucci et al. 2007; PG~1416-129, Porquet et al. 2007;
NGC~4051, Ponti et al. 2006; NGC~3516, Turner et al. 2005; NGC~3227, Gondoin et al. 2003; 1H~0419-577, Pounds et al. 2004); 
all of these sources are characterized by puzzling spectral and temporal behaviors.
In general, the rather flat spectra of type 1 AGN can be ascribed to either the presence of 
a warm and/or complex absorber, or alternatively to a reflection bump as all of these components conspire to 
flatten the primary continuum. Another possible explanation for the detection of many flat spectrum sources 
could be the hard X-ray selection of our sample, i.e. AGN  with a flat $\Gamma$ and a significant 
reflection component (when present) are more easily detected at energies $>$ 10 keV than objects with a 
steep continuum and no reflection. However, the average photon index found in hard X-rays surveys 
is $\Gamma$ $\sim$ 2.1 (see Beckmann et al. 2006, Deluit \& Courvoisier 2003,
Zdziarski et al. 1995) and indeed, if we fit the {\it INTEGRAL} data alone with a simple power-law,
the average photon index turns out to be $\Gamma$ = 2.08 $\pm$ 0.18. Furthermore,
objects with a flat primary continuum are not characterized by a particularly 
high reflection component (in fact $R$ tends to be low in these objects).
Further 2-10 keV data possibly on a complete sample of hard X-ray selected AGN may  
help understanding if these 4 AGN are peculiar and rare objects or instead belong to a typical 
and more abundant than previously thought population.

\begin{figure*}[!ht]
\begin{center}
\parbox{16cm}{
\includegraphics[width=0.45\textwidth,height=0.3\textheight]{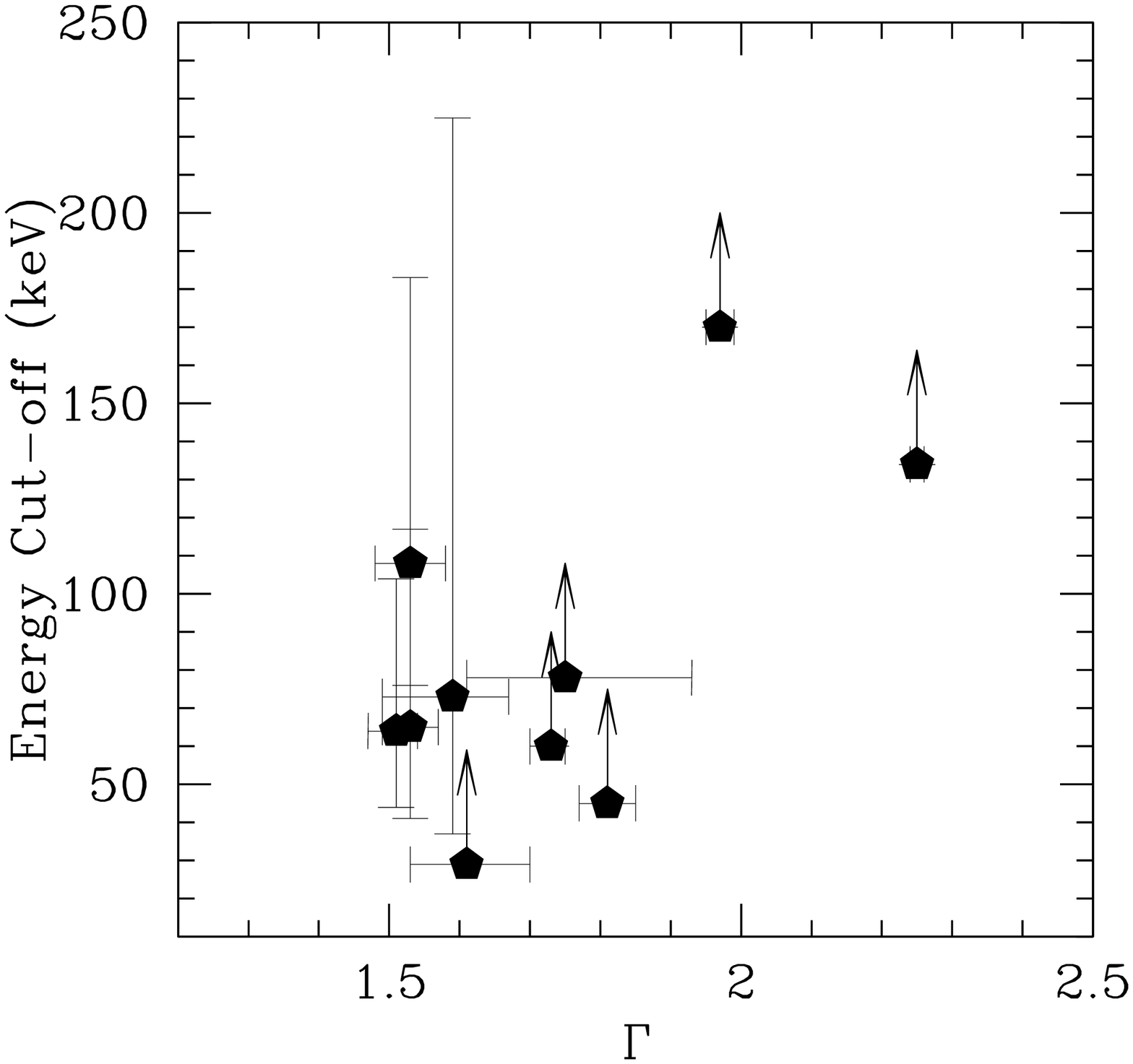}
\includegraphics[width=0.45\textwidth,height=0.3\textheight]{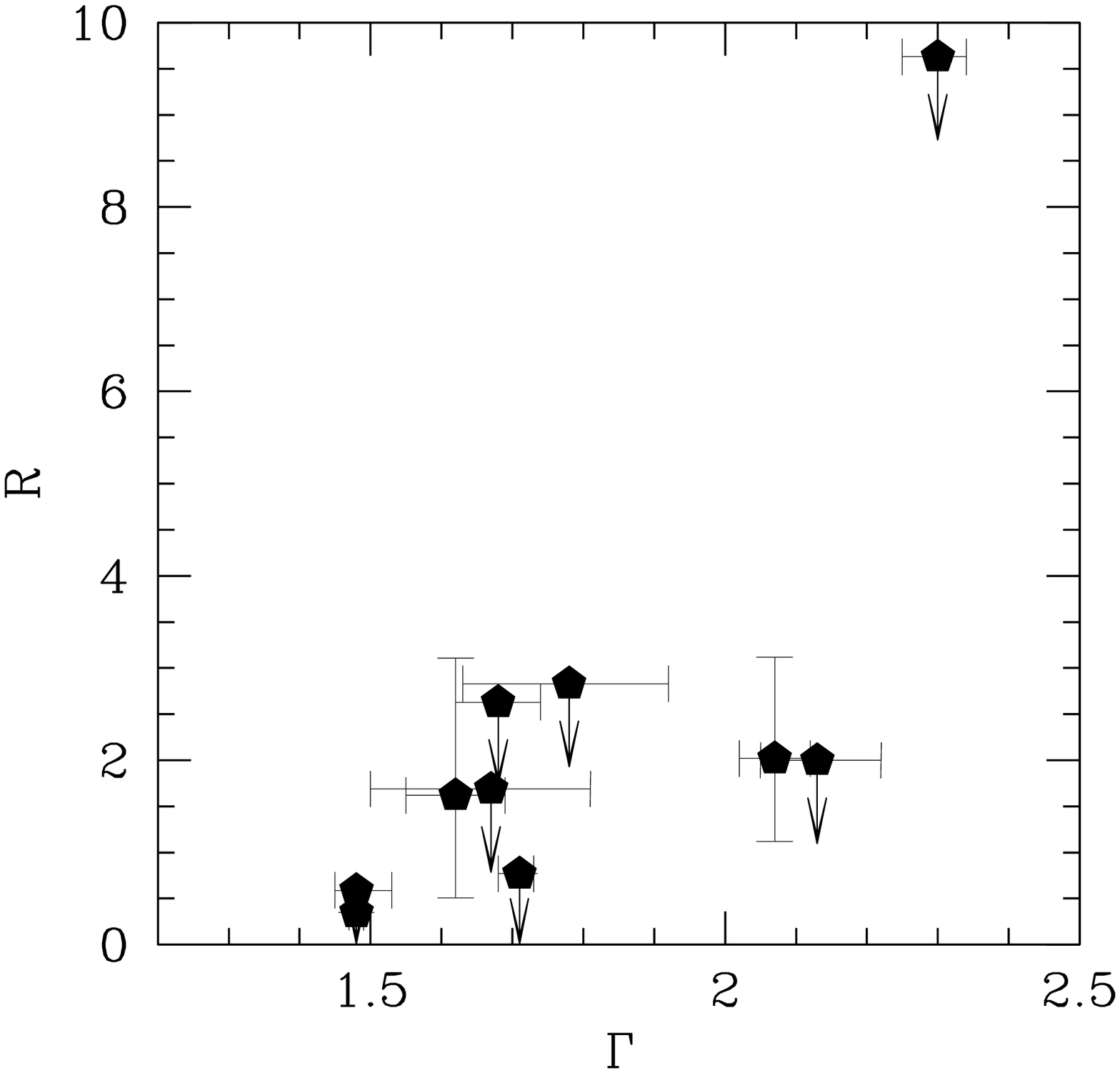}}
\caption{Left panel: Energy cut-off (keV) versus $\Gamma$. Right panel: Reflection component $R$ versus $\Gamma$.}
\label{figure=refl}
\end{center}
\end{figure*}

The broad-band spectral analysis allowed us to constrain the high energy cut-off in
four out of nine AGN in the sample and for all of them
the cut-off energy is found to be below 150 keV. This is at odds with 
previous results which located the cut-off energy, on average, at $\sim$ 200 keV for the type 1 Seyfert 
galaxies observed with {\it BeppoSAX} (Perola et al. 2002, Malizia et al. 2003b, Dadina 2008). 
Gondek et al. (1996), who analyzed the average 1-500 keV
spectrum of type 1 AGN observed with {\it EXOSAT}, {\it Ginga}, {\it HEAO-1} and {\it CGRO} OSSE,
found a cut-off energy $>$ 500 keV; similarly a study by Zdziarski et al. (1995) of broad-band 
{\it Ginga} and {\it CGRO} OSSE spectra of a large sample of Seyferts provided E$_{cut-off}$ $>$ 250 keV.
Moreover, only 6 of the 36 AGN detected by {\it INTEGRAL} analyzed by Beckmann et al. (2006) 
and Sazonov et al. (2004) require a cut-off at energies below 200 keV.
On the other hand, there are spurious examples in the literature of type 1 Seyferts with
low values of the high energy cut-off (e.g., Molina et al. 2006, Perola et al. 2002).
Also the analysis of the average {\it BeppoSAX} PDS spectra by Deluit \& Courvoisier (2003) shows that the lower limit
obtained for type 1 Seyfert galaxies is fully consistent with our results (E$_{cut-off}$ $>$ 63 keV). This is also
consistent with recent results by  Ajello et al. (2008) who found a mean cut-off energy at $\sim$ 100 keV
in Seyfert 1 galaxies observed by {\it Swift}/BAT. 
A positive correlation between the cut-off energy versus the photon index 
was previously found using {\it BeppoSAX} data, first by Piro (1999),
then by Petrucci et al. (2001) and finally confirmed by Perola et al. (2002). 
In Fig.~\ref{figure=refl} (left panel) we plot the cut-off energy (keV) versus 
the photon index $\Gamma$ (data taken from Table~\ref{table=cut}). 
Given the poor statistical quality of our data, we cannot draw any conclusion
on the correlation between these two parameters. However, no evidence
of a correlation is found with the present data (censored data are not included in the analysis),
although the four sources in which the cut-off energy has been constrained
are those showing the flattest photon indeces. We point out that
the known interdependence of the parameters in the {\it pexrav} model
( i.e., the cut-off energy is a variable in the fit which is strongly dependent
on $\Gamma$ and $R$) has to be taken into account when these correlations are
discussed.
 
\subsection{The reflection component and the Fe lines}

A correlation between the reflection parameter $R$ and the photon index 
has been claimed by Zdziarski et al. (1999) from the study of a large number 
of {\it Ginga} observations of Seyfert galaxies and Galactic black holes. However, the
validity of this correlation is under debate, since $R$ and $\Gamma$ are strongly 
correlated in the fitting procedure (Vaughan \& Edelson
2001) and subsequent works did not confirm the presence of a relation between 
these two parameters (Petrucci et al. 2001, Perola et al. 2002). 
Recently, Mattson, Weaver \& Reynolds (2007)
have shown that the strong correlation found between $R$ and $\Gamma$
in their sample of type 1 and type 1.2 Seyfert galaxies observed with {\it RXTE},
is likely to be an artefact of modeling degeneracy. 
The reflection parameter in our sample analysis is constrained
only in IGR~J16482-3036 and IGR~J17418-1212. Therefore, the poor statistics 
and the large number of upper limits prevent us from drawing strong
conclusions on this point. The $R$ versus $\Gamma$ (data taken from Table~\ref{table=pex}) plot is 
shown in Fig.~\ref{figure=refl}, right panel: stronger reflection is indeed measured 
in steeper spectrum sources, but when upper limits are considered this effect is not so obvious in our data.

IGR~J16482-3036, IGR~J17418-1212 and possibly IGR~J16558-5203 show a
reflection component $R$ $>$ 1. These high values of $R$ have been
found in previous studies using {\it ASCA} and {\it BeppoSAX} data
(Cappi et al. 1996, Dadina 2008) and, more recently, with the {\it Suzaku} 
satellite (Miniutti et al. 2007, Comastri et al. 2007).
Such strong reflection might be present when more primary X-ray
radiation is emitted toward the reflector than toward the observer as
is possible in the case of strongly variable nuclear emission or when
there is a time delay between the underlying continuum and the
reflected component, caused by a large distance between the reflecting
material and the primary source (Malzac \& Petrucci 2002). Another
explanation might be a peculiar geometry (Malzac et al. 2001;
Malzac 2001) or general relativistic light bending effects (Fabian et
al 2004; Miniutti \& Fabian 2004; Fabian et al 2005). Recently, Gandhi
et al. (2007) have shown that, in the synthesis models of the X-ray background spectrum,
a significant fraction of this type of source is needed when light bending effects are taken into account.

Under the hypothesis that the line emission is entirely associated
with optically thick material, the equivalent width of the Fe line is
expected to correlate linearly with the value of $R$ obtained for an
arbitrarily fixed inclination angle. To a first approximation, an
equivalent width of 140 eV is predicted if $R$ $=$ 1 for a
cold, face-on disk and an incident power-law with $\Gamma$ $\sim$ 2
(George \& Fabian 1991). In the case of IGR~J16482-3036, the
equivalent width (104$^{+25}_{-82}$ eV), measured using the model in
Table~\ref{table=pex}) is consistent with the observed reflection
fraction; while in the case of IGR~J17418-1212 it is very modest
(41$^{+26}_{-26}$ eV). Some scatter in the equivalent width could be
introduced by a variance in the iron abundance which we do not take
into account in our analysis. On the other hand, this discrepancy can
be explained in terms of a possible anisotropy of the source of seed
photons which might modify the resulting spectrum (Petrucci et
al. 2001, Merloni et al. 2006).

For a given $R$, the equivalent width may also differ according to the value of $\Gamma$.
The simulations of George \& Fabian (1991) showed that
the equivalent width of the iron line should decrease as the
spectrum softens, given the presence of fewer photons with energies
above the iron photoionization threshold. Mattson, Weaver \& Reynolds (2007) 
further indicated that the correlation between these two parameters is not so simple. In fact, they observed
a positive trend which reverses at $\Gamma$ $\sim$ 2, i.e., the equivalent width versus $\Gamma$ 
plot shows a correlation for $\Gamma$ $<$ 2 and an anticorrelation for $\Gamma$ $>$ 2.
Interestingly, the two sources with a constrained value of $R$ reflect
this behavior: IGR~J16482-3036, for which the $R$ value is in agreement with the equivalent width of
the line, has a $\Gamma$ = 1.62 $\pm$ 0.07, while IGR~J17418-1212, in which a low equivalent 
width of the line is found, has $\Gamma$ = 2.07 $\pm$ 0.05.
Broad-band simultaneous observations, such as those performed by {\it Suzaku}, are highly recommended
in order to confirm the presence of a strong reflection component in these objects.

\section{Conclusions}

The IBIS instrument on board {\it INTEGRAL} has allowed the detection
of new bright Seyfert galaxies. Here, we have presented the broad-band
spectral analysis of nine type 1 AGN, 
eight of them never observed before below 10 keV with the high sensitivity of {\it XMM-Newton}.
The simultaneous fitting 
of the EPIC/pn (or MOS) {\it XMM-Newton} and IBIS {\it INTEGRAL} 
data allowed a description of the continuum in terms of an absorbed power-law,
plus a thermal soft component and an FeK emission line. In addition to this
simple description of the continuum, we have found
several additional components, such as partial covering absorption, Gaussian components and/or absorption edges.
We also checked for the presence of an exponential high energy cut-off and a Compton reflection component.
Bearing in mind the limitations of our broad-band analysis, mainly due to the non-simultaneity of the 
{\it XMM-Newton} and {\it INTEGRAL} observations, we summarize our findings as follows:

\begin{itemize}

\item Intrinsic fully-covering absorption has been found in five sources of the sample.
Partial covering absorption is required in 4U~1344-60 and IGR~J16558-5203.
The presence of Compton thin clouds and/or of the outer part of the torus along 
the line of sight to the observer (as predicted in Unified Models) might explain the measured 
column densities as well as the complexity of the absorber. A soft component at
$E < 2 \rm \, keV$ is present in more than half of our sources.

\item A power-law model provides a good approximation of the primary continuum of the AGN presented here and  
in the cases of LEDA~168563, IGR~J07597-3842, IGR~J16482-3036 and IGR~J18027-1455, 
a high energy exponential cut-off is significantly required by the data. 
The introduction of a reflection component is statistically significant only for IGR~J17418-1212.

\item When we consider the best--fit model for each source, the mean photon index 
found is $< \Gamma >$ = 1.73 $\pm$ 0.24, consistent with the mean values 
commonly found in AGN. However, the photon index distribution has a peak at 
$\Gamma$ $\sim$ 1.5 and a tail towards steeper indeces.

\item When constrained, the cut-off energy is located below 150 keV and no evidence for a correlation
between spectral slope and the high energy cut-off 
is found in our data, at odds with results in the literature (Piro 1999, 
Petrucci et al. 2001, Perola et al. 2002). However, the limited
statistics prevent us from drawing firm conclusions on this point.

\item The reflection parameter $R$ has been constrained in only two
  sources, namely IGR~J16482-3036 and IGR~J17418-1212; in both of them
  the reflection fraction is found to be high, i.e. $R$ $>$
  1. According to current theoretical models, the equivalent width of
  the Fe line should correlate linearly with the value of $R$. This
  correspondence is observed in IGR~J16482-3036, but not in the case
  of IGR~J17418-1212. Variance in the iron abundance and/or possible
  anisotropy of the source of seed photons could be responsible for
  the observed mismatch, as well as spectral and flux variability of
  the sources.

\end{itemize}

\begin{acknowledgements}

We thank the anonymous referee for constructive and valuable comments
which improved our manuscript. The results reported here are based on observations obtained
with ESA science missions {\it XMM-Newton} and {\it INTEGRAL}.
The team acknowledges support by ASI-INAF I/023/05/0 grants.
F.P. acknowledges support by a "Juan de la Cierva'' fellowship and 
the Spanish Ministry of Education and Science, under project ESP2006-13608.

\end{acknowledgements}

\end{document}